\magnification=1200

\headline{\ifnum\pageno=1 \nopagenumbers
\else \hss\number \pageno \fi}

\overfullrule=0pt
\footline={\hfil}
\font\boldgreek=cmmib10
\textfont9=\boldgreek
\mathchardef\mypsi="0920

\mathchardef\myphi="091E

\def\lsim{\raise0.3ex\hbox{$<$\kern-0.75em\raise-1.1ex\hbox{$\sim$}}}
\def\gsim{\raise0.3ex\hbox{$>$\kern-0.75em\raise-1.1ex\hbox{$\sim$}}}
\baselineskip=10pt
\vbox to 1,5truecm{}
\parskip=0.2truecm
\centerline{\bf Theoretical Description of the HERA Data on $\bf F_2$} 
\smallskip
\centerline{\bf at Low $\bf Q^2$}\bigskip

\bigskip 
\centerline{\bf  A.B. Kaidalov}
\smallskip
\centerline{ITEP, B. Cheremushkinskaya 25}
\centerline{117259 Moscow, Russia}
\medskip
\centerline{\bf  C. Merino}
\smallskip
\centerline{Dpto. de F\'\i sica de Part\'\i culas}
\centerline{Universidade de Santiago de Compostela}
\centerline{E-15706 Santiago de Compostela, Spain}
\bigskip \bigskip \bigskip\baselineskip=20pt

\noindent
${\bf Abstract}$ \par
It is shown that the CKMT model for the nucleon structure function $F_2$ gives
a good description of the recent HERA data at low and moderate $Q^2$. Also
the fit to the same data obtained with a modified version of the model in which
a logarithmic dependence on $Q^2$ has been included, is presented. For moderate
values of $Q^2$, in the
current available range of $x$, the first parametrization leads to a better
description of the data. 
\par

\vbox to 4 truecm{}

\noindent June 1998
\vfill \supereject
\noindent {\bf 1. \underbar{Introduction}} \par \vskip 5 truemm
The CKMT model [1] for the parametrization of the nucleon structure function
$F_2$ is a theoretical model based on Regge theory which provides a consistent
formulation of this function in the region of low $Q^2$, and that can therefore
be used as a safe and theoretically justified initial condition in the
perturbative QCD evolution equation, to obtain the structure function at larger
values of $Q^2$. Thus, the CKMT model, which gives a good description [1] of all
the pre-HERA measurements [2] and of the first small-$x$ experimental data from
HERA [3] on $F_2(x,Q^2)$ and $\sigma^{tot}_{\gamma p}(\nu)$, is a useful tool
to reach deeper insight on the crucial interplay between soft (low $Q^2$) and
hard (high $Q^2$) physics that now for the first time is being studied in the
two small-$x$ experiments (H1 and Zeus) at HERA.

In this Letter, we present the description that the CKMT model provides for the
more recent published data on $F_2$ at low $Q^2$ by both HERA experiments [4,5],
which should contribute to determine the  initial condition for the 
perturbative QCD evolution equation. Also we present, the description of
 the same 
experimental data
 by a modified version of the CKMT model in which  a
logarithmic dependence
on $Q^2$ asymptotically predicted by the perturbative QCD,
has been included. Even though both descriptions lead to good
descriptions of experimental data in  the present
experimental range of $x$ for moderate values of $Q^2$, the fit obtained with
the non-modified version of the CKMT model shows a better agreement and
has more natural values of parameters .
\par \vskip 5 truemm

\noindent {\bf 2. \underbar{The model}} \par \vskip 5 truemm

The CKMT model [1,6] proposes, taking into account what we know from Regge
theory and hadronic interactions, for the nucleon structure functions
$$ F_2(x,Q^2) = F_S(x,Q^2) + F_{NS}(x,Q^2),
\eqno(1)$$
\noindent 
the following parametrization of its two terms in
the region of small and moderate $Q^2$. For the singlet term, corresponding to
the Pomeron contribution:
$$ F_S(x,Q^2) = A\cdot x^{-\Delta(Q^2)}\cdot(1-x)^{n(Q^2)+4}
\cdot\left({Q^2\over Q^2+a}\right)^{1+\Delta(Q^2)},
\eqno(2)$$
\noindent
which $x$$\rightarrow$0 
behavior is determined by an effective intercept of the Pomeron, $\Delta$, 
which takes into account Pomeron cuts and, therefore (and this is one of the
main points of the model) it depends on $Q^2$. This dependence was 
parametrized in [1]
as :
$$ \Delta (Q^2) = \Delta_0\cdot\left(1+{\Delta_1\cdot Q^2
\over Q^2+\Delta_2}\right). 
\eqno(3)$$
\noindent
Thus, for low values of $Q^2$ (large cuts), $\Delta$ is close
to the effective value found from analysis of hadronic total cross sections
($\Delta$$\sim$0.08), while for high values of $Q^2$ (small cuts),
$\Delta$ takes the bare Pomeron value, 
$\Delta$$\sim$0.2-0.25. The
parametrization for the non-singlet term, which corresponds to the secondary
reggeon (f, $A_2$) contribution, is:   
$$F_{NS}(x,Q^2) = B\cdot x^{1-\alpha_R}\cdot(1-x)^{n(Q^2)}
\cdot\left({Q^2\over Q^2+b}\right)^{\alpha_R},
\eqno(4)$$
\noindent
where the $x$$\rightarrow$0 behavior is determined by the secondary
reggeon intercept $\alpha_R$, which should be in the range $\alpha_R$=0.4-0.5.
The valence quark contribution can be separated into the contribution of the u
and d valence quarks by replacing 
$$ B\cdot (1-x)^{n(Q^2)}\rightarrow  B_u\cdot (1-x)^{n(Q^2)} + B_d\cdot
(1-x)^{n(Q^2)+1}, 
\eqno(5)$$
\noindent
and the normalization condition for valence quarks fixes $B_u$ and $B_d$ at one
given value, $Q_0^2$, of $Q^2$ (we use $Q_0^2=2.GeV^2$ in our calculations).
For both the singlet and the non-singlet terms, the behavior when 
x$\rightarrow$1 is
controlled by $n(Q^2)$, with $n(Q^2)$ being
$$ n(Q^2) = {3\over2}\cdot\left(1+{ Q^2
\over Q^2+c}\right),
\eqno(6)$$
\noindent
so that, for $Q^2$=0, valence quark distributions have the same power, given by
Regge intercepts, as in Dual
Parton Model [7], $n$(0)=$\alpha_R$(0)$-$$\alpha_N$(0)$\sim$ 3/2, and the behaviour of
$n(Q^2)$ for large $Q^2$ is chosen to coincide with dimensional counting rules
 [8].

The total cross section for real ($Q^2$=0) photons can be obtained from the
structure function $F_2$ using the following relation: 
$$ \sigma^{tot}_{\gamma p}(\nu) = \left[{4\pi^2\alpha_{EM}\over Q^2}
\cdot F_2(x,Q^2)\right]_{Q^2=0}.
\eqno(7)$$
\noindent
The proper $F_2(x,Q^2)$$\sim$$Q^2$ 
behavior at
$Q^2$$\rightarrow$0, is fulfilled in the model
 due to the last factors in equations (2) and (4), which 
were
taken in the same form as in references [9] and [10]. Thus, the following 
parametrization of
$\sigma^{tot}_{\gamma p}(\nu)$ takes place in the CKMT model:
$$ \sigma^{tot}_{\gamma p}(\nu) = 4\pi^2\alpha_{EM}
\cdot\left(A\cdot a^{-1-\Delta_0}\cdot(2m\nu)^{\Delta_0}
+(B_u+B_d)\cdot b^{-\alpha_R}\cdot(2m\nu)^{\alpha_R-1}\right).
\eqno(8)$$
\noindent
Therefore, the CKMT parametrizes both the nucleon
structure functions and $\gamma p$ total cross-section with a 8-parameter
function.  Besides the normalizations, 4 of the parameters appear in (8), so
the proton structure function contains only 4 extra parameters. Although the
parameters are not completely free and they are correlated,
 some theoretical uncertainty
still exists in the determination of their exact values. To solve this
uncertainty, comparison with experimental data is needed. From this comparison
with experiments, a good description of all available experimental data on
total cross-section for real photons and nucleon structure functions from 
former pre-HERA experiments~[2], and from the first HERA measurements~[3] was 
obtained
in~[1] by using the CKMT model with the values for the different parameters
which are listed in Table~1, (a), and which were found without including any 
HERA
data in the fit. It has also to be reminded that the good accuracy of the fit 
is not very sensitive to changes in the values of the parameters inside the
ranges allowed by theory. Moreover, the model provides [11,12] in addition 
reasonable descriptions of the HERA data on diffraction [13], through the 
parametrization of the
Pomeron structure function, 
 and of the available 
experimental data on nuclear shadowing [14], by using this parametrization of
the Pomeron structure function in the frame of the Gribov theory [15].

\par 
\vskip 5 truemm
\vfill\supereject

\noindent {\bf 3. \underbar{Description of the HERA data on $F_2$ at low $Q^2$}}\par \vskip 5 truemm
In the former fit of reference [1], the lack of experimental data at low and
moderate $Q^2$ limited the accuracy in the determination of the values of the
parameters in the model. Now, the publication of the new experimental
data [4,5] on $F_2$ from HERA at low and
moderate $Q^2$ gives us the opportunity to include in the fit of the parameters
of the model experimental points from HERA experiments from the kinematical
region where the parametrization should give a good description without need of
any perturbative QCD evolution.\par

Thus, we proceed as one had done in [1], but by adding the above mentioned
experimental data on $F_2$ from H1 and Zeus at low and moderate $Q^2$, to
those~[2] from NMC and E665 collaborations, and to data on
cross-sections for real photoproduction, into a global fit which allows the test
of the model in larger regions of $x$ and $Q^2$. We take as initial condition
for the values of the different parameters
those obtained in the previous fit~[1].
The result of the new common fit to $\sigma^{tot}_{\gamma p}$
and $F_2$ is presented in Figs. 1 and 2, and the final values of the parameters
can be found in Table 1, (b). Now, the parameter $\Delta_1$, which was fixed
to $\Delta_1$=2.0 in former fits, (a), where also the
notation $d$ had been used~[1] for the parameter $\Delta_2$, has been left 
free, 
and, 
since the 
present global fit 
turns out 
to be not very sensitive to changes in the original values of the parameters
$c$ and $\alpha_R$, we keep them fixed. As it can be seen in the figures, the
quality of the
description provided by the CKMT model of all the experimental data, and, in
particular, of the new experimental data from HERA is very good, with a value
of $\chi^2/d.o.f.$ for the global fit, $\chi^2/d.o.f.$=106.95/167, where 
the statistical and 
systematic
errors have been treated in quadrature, and where the relative normalization
among all the experimental data sets has been taken equal to 1.\par

Also, since the small-$x$ HERA experiments allow for the first time the 
experimental study of the question of the interplay between soft and hard 
physics, we
have modified our model,  with basically only power dependence
on $Q^2$, to include a logarithmic 
dependence on 
$Q^2$ as the one predicted
asymptotically by perturbative QCD [16]. By doing so, we try to know
whether such a modified version of the model provides a smoother matching in the
description of both regimes, and whether the available experimental data can
distinguish the description obtained with a soft model in which the
$Q^2$-dependence saturates, from that of a model which does not present such a
saturation.\par

To include the logarithmic dependence on $Q^2$ in our model, we take into
account that the behavior of $F_2$ at small $x$ in QCD, is given by the
singularities of the moments of the structure functions~[17], the rightmost
singularity giving the leading behavior. Thus, we introduce in (1) the following factors~[17,18], which correspond to the moments of the structure functions
in the language of the OPE expansion, and that
can be calculated by
making the convolution in rapidity of the hard-upper part with the soft-lower
part of the leptoproduction diagram: 
$$ \left({\alpha_s(Q^2_0)\over\alpha_s(Q^2)}\right)^{d_i(n_i)}, i=S,NS,
\eqno(9)$$
\noindent
where the strong coupling constant is taken as
$$ \alpha_s(Q^2) = {4\pi\over\beta_0\cdot log\left({Q^2+M^2\over 
\Lambda_{QCD}^2}\right)},
\eqno(10)$$
\noindent
with M$\sim$1GeV, a hadronic mass~[19] included in (10) to avoid the 
singularity in
$\alpha_s$ when $Q^2$$\rightarrow \Lambda_{QCD}^2$, $\Lambda_{QCD}$=0.2 GeV, and
$\beta_0$=11$-{2\over3}n_f$ (we use in our calculations a number of flavors, 
$n_f$=3), and where $d_S(n_S)$ and $d_{NS}(n_{NS})$ are  respectively 
proportional to
the largest eigenvalue of the anomalous dimension matrix, and to the anomalous
dimension:
$$ d_S(n_S) \sim {d_0\over4(n_S-1)}-d_1, 
\eqno(11)$$
\noindent
with
$$ d_0 = {48\over\beta_0}, d_1 = {11+{2\over 27}n_f\over\beta_0},
$$
\noindent
and
$$ d_{NS}(n_{NS}) = {16\over33-2n_f}\cdot\left({1\over2n_{NS}(n_{NS}+1)}
+{3\over4}-S_1(n_{NS})\right), 
\eqno(12)$$
\noindent
with
$$ S_1(n_{NS}) = n_{NS}\cdot\sum^{\infty}_{k=1}
{1\over k(k+n_{NS})}.
$$
\par

Thus, (1) is modified in the following 
way:
$$ F_2(x,Q^2) = \left({\alpha_s(Q^2_0)\over\alpha_s(Q^2)}\right)^{d_S(n_S)}
\cdot F_S(x,Q^2) +
\left({\alpha_s(Q^2_0)\over\alpha_s(Q^2)}\right)^{d_{NS}(n_{NS})}
\cdot F_{NS}(x,Q^2).
\eqno(13)$$
\noindent

The exponents of the new factors in (13), $d_S(n_S)$ and $d_{NS}(n_{NS})$,
give us the singularities in $n_i$, $i$=$S$,$NS$, of the momenta, which, as we 
mentioned
above, control the QCD small-$x$ behavior of $F_2$. Therefore, in our model,
these exponents   
have to be evaluated, see (2) and (4), at 
$n_S$=1+$\Delta(Q^2$$\rightarrow$$\infty$)=
1+$\Delta_0$(1+$\Delta_1$) and $n_{NS}$=$\alpha_R$, respectively. 
We consider in our expressions only the
LO behavior, and again we can use the relation (3) to write the total 
cross-section for real
photoproduction from (13).\par

At this point, two main differences between our approach and the one of
reference~[18] which also provides good fits of the experimental data, can be 
mentioned. First, while in our model just one Pomeron with $Q^2$-dependent
intercept is present,
in the approach of ref. [18] 
 two components,- a constant soft Pomeron and a hard Pomeron
with very large intercept $\Delta$$\sim0.5$,  
 are used. 
Secondly, the proper $Q^2$
behavior (see also~[10]) of $F_2$ 
for $Q^2\rightarrow$0 is provided in a natural way. 
So when
the logarithmic dependence on $Q^2$ is included in the model, a justified 
choice~[19] of the effective 
mass $M^2$ in $\alpha_s(Q^2)$ (see equation (10)) provides a non-singular
behavior of $\alpha_s(Q^2)$ throughout the $Q^2$ range.
 In~[18], on the contrary, one particular value of 
$M^2$=$\Lambda_{QCD}^2$ has to be taken in (10) in order to get the required 
behavior  
of $F_2$ for $Q^2\rightarrow$0, thus resulting 
in a
singular behavior $\alpha_s(Q^2)$$\rightarrow$$\infty$, when 
$Q^2$$\rightarrow$0.\par

Then, we use this modified version of the CKMT parametrization of $F_2$ to repeat the fit of the same experimental data, including the HERA data on $F_2$ at
small and moderate $Q^2$. As starting point for
the QCD evolution, we take the same value $Q^2_0$=2.$GeV^2$ we use to fix the
normalization of the valence component.
The result of this second fit is also presented in
Figures 1 and 2, and the final values of the parameters in the model are given
in Table 1, (c). Now, only the parameter $c$ has been kept fixed to its
original value in (a).\par

As it can
be seen in the figures, the quality of this second fit is also reasonable,
although the value of $\chi^2/d.o.f.$ ($\chi^2/d.o.f.$=453.19/167), is now 
appreciably higher than in the fit obtained
with the non-modified version of the CKMT model.\par
 
One common feature of the two fits is that both give higher 
values 
of
the total cross-section for real photons than the experimental ones, in
the region of large W where the experimental error bars are  large. More
accurate measurements in this region should clarify this point. Concerning the
final values of the parameters, it has to be noticed that $\Delta_0$ takes 
in (b) and (c) similar values,
both slightly larger than the original one in (a), providing a stable 
asymptotic behavior for the effective 
$\Delta$($Q^2$$\rightarrow$$\infty$)$\sim$0.25-0.3
which includes some effects of perturbative QCD evolution.\par
\vskip 5 truemm

\noindent {\bf 4. \underbar{Conclusions}} \par \vskip 5 truemm

In conclusion, the CKMT model for the parametrization of the nucleon
structure functions
provides a very good description of all the
available experimental data on $F_2(x,Q^2)$ at low and moderate $Q^2$, 
including the more 
recent small-$x$ HERA points. Also the fit to the same data obtained with a 
modified version of the model in 
which a
logarithmic
dependence on $Q^2$ is included, has been presented. Eventhough the quality of
this second description is also reasonable, its $\chi^2/d.o.f.$ is appreciably
higher 
than that corresponding to the fit obtained with the non-modified version of
the CKMT model. 
\par \vskip 5 truemm
\vfill \supereject

\noindent {\bf Acknowledgements} \par
We are grateful to A.~Capella, K.G.~Boreskov, G.~Parente,Yu.~A.~Simonov  and 
R.~V\'azquez for 
useful 
discussions. One
of us (C.M.) also wants to thank the ITEP members for their kind hospitality
during the realization of this work.\par \vskip 5 truemm

\centerline{\bf References} \vskip 3 truemm
\item{[1]} A.~Capella, A.B.~Kaidalov, C.~Merino and J.~Tran~Than~Van, Phys. 
Lett. {\bf B337} (1994) 358. 
\item{[2]} New Muon Collaboration, P.~Amaudruz et al., Phys. Lett. {\bf B259} 
(1992) 159.
\item{} E665 Collaboration, M.R.~Adams et al., FERMILAB-Pub 1995/396.
\item{[3]} H1 Collaboration, I.~Abt et al., Nucl. Phys. {\bf B407} (1993) 515.
 \item{} Zeus Collaboration, M.~Derrick et al., Phys. Lett. {\bf B316} (1993) 
412.
\item{[4]} H1 Collaboration, C.~Adloff et al. Nucl. Phys. {\bf B497} (1997) 3.
\item{[5]} Zeus Collaboration, J.~Breitweg et al., Phys. Lett. {\bf B407} 
(1997) 432.
\item{[6]} A.B.~Kaidalov, Survey in High Energy Physics {\bf 9} (1996) 143.
\item{[7]} A.~Capella, U.~Sukhatme, C.-I.~Tan and J.~Tran~Than~Van, Phys.
Rep. {\bf 236} (1994) 225.
\item{[8]} S.~Brodsky, G.~Farrar, Phys. Rev.
Lett. {\bf 31} (1973) 1153.
\item{}  V.~Mateev, R.~Muradyan, and A.~Tavkhelidze,
Nuovo Cim. {\bf 7} (1993) 719.
\item{[9]} H.~Abramowicz, E.M.~Levin, A.~Levy and U.~Mahor, Phys.
Lett. {\bf B269} (1991) 465.
\item{[10]} A.~Donnachie, P.V.~Landshoff, Z. Phys.$-$Particles and Fields
{\bf C61} (1994) 139.
\item{[11]} A.~Capella, A.B.~Kaidalov, C.~Merino and J.~Tran~Than~Van, Phys.
Lett. {\bf B343} (1995) 403.
\item{}  A.~Capella, A.B.~Kaidalov, C.~Merino, D.~Pertermann and J.~Tran~Thanh~
Van, Phys. Rev. {\bf D53} (1996) 2309.
\item{[12]} A.~Capella, A.B.~Kaidalov, C.~Merino, D.~Pertermann and
J.~Tran~Thanh~Van, to be published in European Journal of Physics.
\item{[13]} H1 Collaboration, T.~Ahmed et al., Nucl. Phys. {\bf B429} (1994)
477.
\item{} H1 Collaboration, T.~Ahmed et al., Phys. Lett. {\bf B348} (1995)
681.
\item{} Zeus Collaboration, M.~Derrick et al., Phys. Lett. {\bf B315} (1993)
481; {\bf B332} (1994) 228; {\bf B338} (1994) 477.
\item{} Zeus Collaboration, B.~Foster, in Proceedings of the Workshop on DIS
and QCD, Paris (1995). 
\item{[14]}  M. Arneodo, Phys. Reports {\bf 240} (1994) 301 (and references
therein).
\item{[15]} V. N. Gribov, ZhETF {\bf 57} (1967) 654 [Sov. Phys. JETP {\bf 26}
           (1968) 14].
\item{[16]} Yu.L.~Dokshitzer, D.I.~Dyakonov and S.I.~Troyan, Phys. Rep. {\bf
58 (5)} (1980) 269.
\item{[17]} A.~de~R\'ujula et al., Phys. Rev. {\bf D10} (1974) 1649.
\item{[18]} F.~Barreiro, C.~L\'opez and F.J.~Yndur\'ain, Z. Phys.$-$Particles
and Fields {\bf C72}
(1996) 561.
\item{} K.~Adel, F.~Barreiro and F.J.~Yndur\'ain, Nucl. Phys. {\bf B495} (1997)
221.
\item{[19]} Yu.~A.~Simonov, Yad. Fiz. {\bf 58} (1995) 113.
\item{[20]} D.O.~Caldwell et al., Phys. Rev. Lett. {\bf 40} (1978) 1222.
\item{[21]} Zeus Collaboration, M.~Derrick et al., Phys. Lett. {\bf B293} 
(1992) 465.
\item{} Zeus Collaboration, M.~Derrick et al., Z. Phys.$-$Particles and Fields 
{\bf C63} (1994) 391.
\item{[22]} H1 Collaboration, S.~Aid et al., Z. Phys.$-$Particles and Fields 
{\bf C69} (1995) 27.
\vfill \supereject

\centerline{
\vbox {\offinterlineskip
\hrule
\halign{&\vrule#&
\strut\quad\hfil#\quad\cr
height2pt&\omit&&\omit&&\omit&&\omit&\cr
&CKMT model\hfil&&{\bf (a)\hfil}&&{\bf (b)\hfil}&&{\bf (c)\hfil}&\cr
height2pt&\omit&&\omit&&\omit&&\omit&\cr  
\noalign{\hrule}
height2pt&\omit&&\omit&&\omit&&\omit&\cr
&A\hfil&&0.1502\hfil&&0.1301\hfil&&0.1188\hfil&\cr
height2pt&\omit&&\omit&&\omit&&\omit&\cr
\noalign{\hrule}
height2pt&\omit&&\omit&&\omit&&\omit&\cr
&a\hfil&&0.2631\hfil&&0.2628\hfil&&0.07939\hfil&\cr
height2pt&\omit&&\omit&&\omit&&\omit&\cr
\noalign{\hrule}
height2pt&\omit&&\omit&&\omit&&\omit&\cr
&$\Delta_0$\hfil&&0.07684\hfil&&0.09663\hfil&&0.1019\hfil&\cr
height2pt&\omit&&\omit&&\omit&&\omit&\cr
\noalign{\hrule}
height2pt&\omit&&\omit&&\omit&&\omit&\cr
&$\Delta_1$\hfil&&2.0\hfil&&1.9533\hfil&&1.2527\hfil&\cr
height2pt&\omit&&\omit&&\omit&&\omit&\cr
\noalign{\hrule}
height2pt&\omit&&\omit&&\omit&&\omit&\cr
&$\Delta_2$\hfil&&1.1170\hfil&&1.1606\hfil&&0.1258\hfil&\cr
height2pt&\omit&&\omit&&\omit&&\omit&\cr
\noalign{\hrule}
height2pt&\omit&&\omit&&\omit&&\omit&\cr
&c\hfil&&3.5489\hfil&&3.5489 (fixed)\hfil&&3.5489 (fixed)\hfil&\cr
height2pt&\omit&&\omit&&\omit&&\omit&\cr
\noalign{\hrule}
height2pt&\omit&&\omit&&\omit&&\omit&\cr
&b\hfil&&0.6452\hfil&&0.3840\hfil&&0.3194\hfil&\cr
height2pt&\omit&&\omit&&\omit&&\omit&\cr
\noalign{\hrule}
height2pt&\omit&&\omit&&\omit&&\omit&\cr
&$\alpha_R$\hfil&&0.4150\hfil&&0.4150 (fixed)\hfil&&0.5872\hfil&\cr
height2pt&\omit&&\omit&&\omit&&\omit&\cr}
\hrule}}

{\parindent= 2 truecm
\item{\bf Table 1 :} Values of the parameters in the CKMT model obtained in
former fits, (a), in the fit in which also the low $Q^2$ HERA data~[4,5]
have been included, (b), and in the fit to the same data obtained with the
modified version of the CKMT model in which a logarithmic dependence of $F_2$
on $Q^2$ has been taken into account, (c). All dimensional parameters are given in
$GeV^2$. The
valence counting rules provide the following values of $B_u$ and $B_d$, for the
proton case, when fixing their normalization at $Q_0^2$=2.$GeV^2$: (a)
$B_u$=1.2064, $B_d$=0.1798; (b) $B_u$=1.1555, $B_d$=0.1722; (c) $B_u$=0.6862,
$B_d$=0.09742. In
previous fits, (a), the
parameter $\Delta_1$ had been fixed to a value $\Delta_1$=2.}

\vfill \supereject
\centerline{\bf Figure Captions}
\vskip 3 truemm
{\parindent= 2 truecm
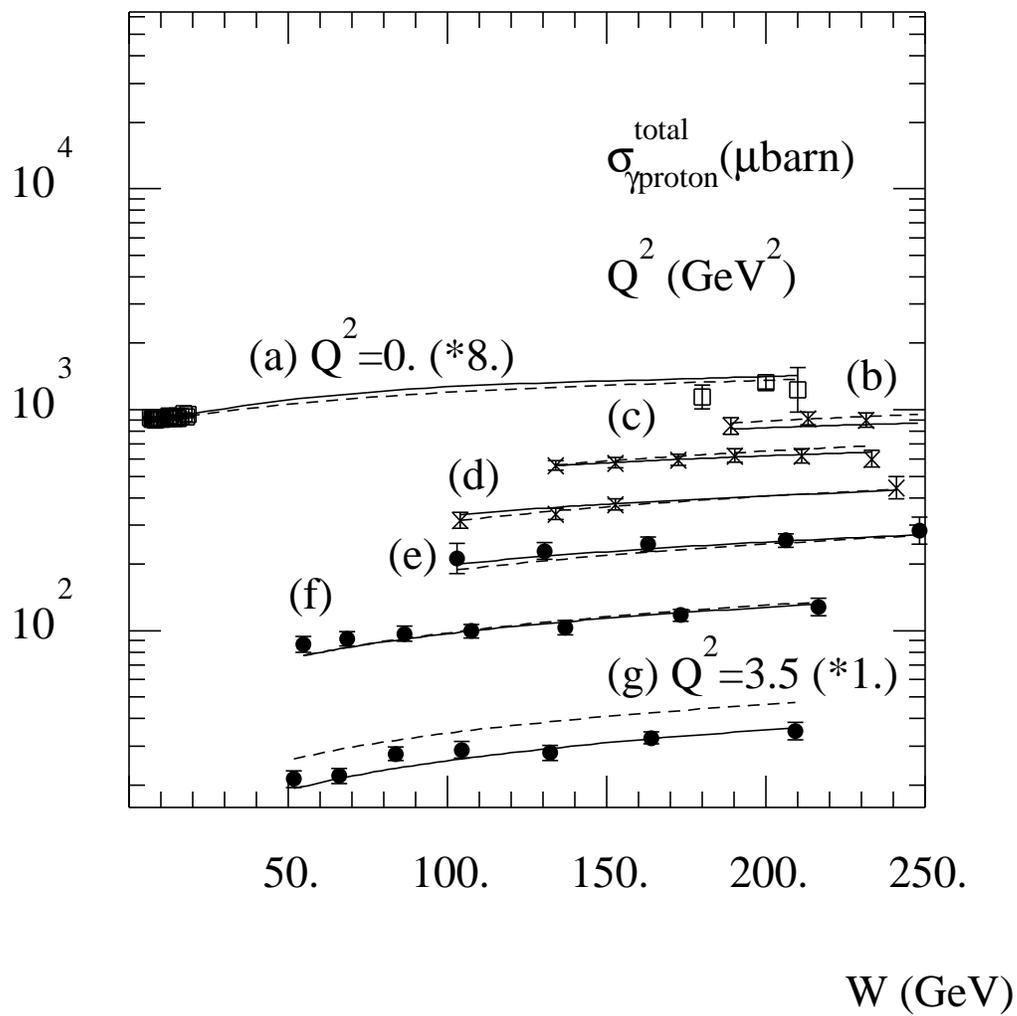
\item{\bf Fig. 1 :} $\sigma^{total}_{\gamma p}$ and
$\sigma^{total}_{\gamma^* p}$
(in $\mu$barns) vs W (in GeV) for different values of $Q^2$. Theoretical fits
have been
obtained with the CKMT
model (full line) and the modified version of the CKMT model (dashed line).
Points at (a), $Q^2$=0.$GeV^2$ (*8.), (b), $Q^2$=0.15$GeV^2$ (*6.),
(c), $Q^2$=0.25$GeV^2$ (*5.), (d), $Q^2$=0.5$GeV^2$ (*4.),
(e), $Q^2$=0.8$GeV^2$ (*3.), (f), $Q^2$=1.5$GeV^2$ (*2.),
and (g), $Q^2$=3.5$GeV^2$ (*1.).
Experimental points for $F_2$ ($\sigma^{total}_{\gamma^* p}$) are from
references~[4],~(black circles), and~[5],~(crosses), and  experimental data on
$\sigma^{total}_{\gamma p}$ are
from references~[20,21,22].
\vskip 3 truemm

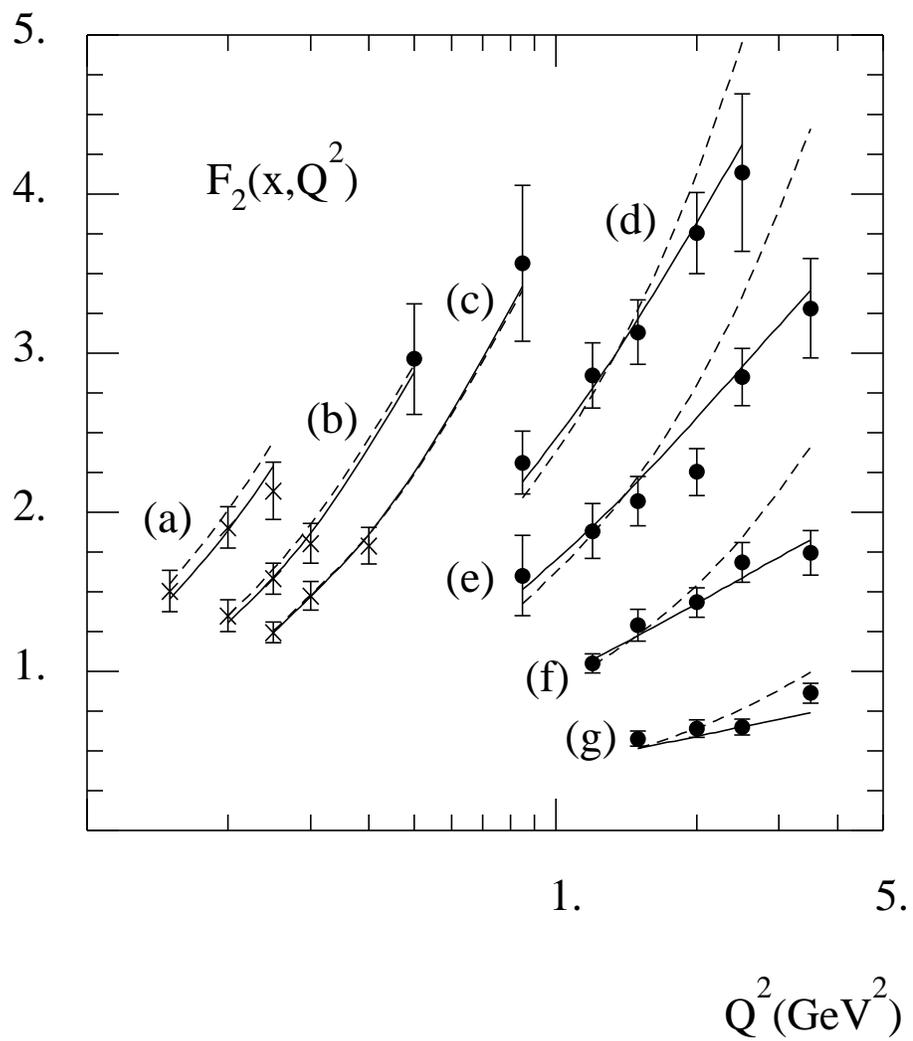
\item{\bf Fig. 2 :} $F_2$($x$,$Q^2$) vs $Q^2$ (in $GeV^2$) for different values
of x. Theoretical fits have been 
obtained with the CKMT
model (full line) and the modified version of the CKMT model (dashed line).
Experimental points at (a), from left to right, x=0.42$\cdot$$10^{-5}$, 
x=0.44$\cdot$$10^{-5}$, 
and x=0.46$\cdot$$10^{-5}$~(*8.); 
(b), from left to right,  x=0.85$\cdot$$10^{-5}$, x=0.84$\cdot$$10^{-5}$,
 x=0.83$\cdot$$10^{-5}$, and x=0.86$\cdot$$10^{-5}$~(*6.);
(c), from left to right, x=0.13$\cdot$$10^{-4}$, and three points at 
x=0.14$\cdot$$10^{-4}$~(*5.); (d),
x=0.5$\cdot$$10^{-4}$~(*4.); (e), x=0.8$\cdot$$10^{-4}$~(*3.); 
(f), x=0.2$\cdot$$10^{-3}$~(*2.); (g), x=0.5$\cdot$$10^{-3}$~(*1.).
Experimental points for $F_2$ are from 
references~[4],~(black circles), and~[5],~(crosses).

\par
}

\bye